\documentclass[manuscript]{acmart}
\usepackage{subfigure}
\usepackage{paralist}
\begin{document}
\title{Effective Communication of Scientific Results}

\author{Jos\'e Nelson Amaral, University of Alberta, Canada}
\maketitle

Communication is essential for the advancement of Science. Technology advances and the proliferation of personal devices have changed the ways in which people communicate in all aspects of life. Scientific communication has also been profoundly affected by such changes, and thus it is important to reflect on effective ways to communicate scientific results to scientists that are flooded with information. This article advocates for receiver-oriented communication in Science, discusses how effective oral presentations should be prepared and delivered, provides advice on the thought process that can lead to scientific papers that communicate effectively, discusses suitable methodology to produce experimental data that is relevant  and offers advice on how to present such data in ways that lead to the formulation of correct claims that are supported by the data.


\section{Receiver-Oriented Communication}

Many years ago I went to a lecture by Preston Manning, a well-known retired politician in Canada, in the then new Science building at the University of Alberta. I expected to hear about geopolitics and major socio-economic issues in Canada. Instead, Manning delivered a lecture on sender-oriented communication vs receiver-oriented communication. While he has since talked about this principle in many contexts, including the failure of elites to communicate with the general public~\cite{ManningGM16}, on that night his focus was to contrast the typical communication styles of an effective politician with that of a successful scientist. A memorable example that he conjured would contrast the politician speaking at the opening of a new child-care centre and the scientist explaining a new method to generate electrical energy. In his telling the scientist would systematically and sequentially go over all the ideas explored  in order to deliver the new generator. She would describe many false starts and explain the way in which obstacles would have been overcome. In contrast the politician would focus on the people around her, talk about the impact of the care centre to families, perhaps tell an anecdote about real people impacted by the centre. Even though the politician may have had many false starts and many obstacles when she tried to get the centre approved and built, when communicating at that moment, she will focus on the people in front of her, on what they are interested to hear, on what matters to them. The scientist, on the other hand, is acting as a sender-oriented communicator. She is focused on the things that matter to her, on the troubles and difficulties that she encountered and how she overcame them. That single lecture changed the way that I see lecturing at the university, advising and coaching students, or speaking at conferences. I recognized that many of us are sender-oriented speakers and that to be better communicators we need to improve our focus on the audience.

A very significant portion of an academic or research scientist time is spent communicating. This communication takes many different formats: lectures, teleconferences, manuscripts, informal discussions. The high-level principles needed to be effective at each  of these interactions are the same: they require a focus on the listeners or the readers. This article discusses several aspects of the communication of Science and also covers some issues related to the experimental methodology used to obtain the scientific results that will be communicated. Some of it will be advice on effective presentations, but it also discusses several aspects of written communication, including the presentation and summarization of data and the writing of claims and conclusions based on the data collected. Section~\ref{sec:perform} provides practical advice on experimental methodology, including the setup required for proper data collection, the need to review experimental frameworks carefully, and a pernicious error in the summarization of results in the computing literature: the incorrect use of the arithmetic average.

This article is a summarization of principles that were developed from reading similar articles, from conversations with graduate students and colleagues over many years and from observing presentations that work and presentations that do not work. Some of the advice on the writing of manuscripts comes from almost twenty years of a tradition in my research group. We have regular round-table discussions of papers. Participants read the paper beforehand and during the discussion I ask one of the students to lead the discussion in a given section of the paper. The student can either explain the section summarizing its main points, or the student may ask questions about anything that the student has not understood in that section. The advantage of this format is that there is not an active speaker and many passive listeners. Students do not know ahead of time which section they may be asked to lead, and therefore all stay attentive throughout the discussion. We have established a tradition that these meetings always start on time and never run pass the 90 minutes time allocated for the meeting. Having a time budget also keeps the discussion focused.

\section{Effective Oral Presentations}

A  student who is tasked to present results of long-term research in a brief presentation often feels that the time is too short and that it is not possible to explain all that they have done. In fact, they should not explain all that they have done. They should only present what matters to the audience. A good way to put the time budget into perspective is to watch a well-made Science-themed show on TV for the amount of time that the presentation is given. And then to reflect on how the TV show was able to present that much information in that amount of time. The answer is that there were no five-bullet text slides in the TV show. Most likely, the presentation was a narrative superposed into videos and images. The combination of the oral presentation with images is often much more effective than the combination of oral presentation with written text. Thus a good presentation starts with the careful preparation of the visual information that will be used to support the presentation and ends with a well planned, properly rehearsed, and well delivered presentation. This section breaks down the process that is required to prepare and deliver an effective presentation. The assumption is that this presentation is in the area of Science and thus it contains greater focus on elements that are typical in a presentation in Science such as block diagrams, graphs, illustrations, equations and, in the context of Computing Science, algorithms and program source code.

\subsection{The Process of Preparing Slides for a Presentation}

A helpful mental framework is to imagine that you are making a short movie instead of an oral presentation. The process of creating a movie includes the generation of ideas and the creation of a storyline. Often this storyline is laid out on a storyboard that is used to decide which scenes will be presented and in which order. Using modern technology, one can start with a set of empty slides and add just a few words to each slide. These words on slides  are similar to the outline of a manuscript and should be chosen to tell the speaker, while organizing the presentation, which are the topics that should be covered, what supporting data needs to be presented, how an argument will be made. In the process of preparing the presentation, a presenter can quickly peruse through these bare slides and reorder them as the mental model of the actual presentation comes into shape. 

Then it is time for the actual making of the film, i.e. to add the graphical elements that will help the listener to understand the presentation. In this process, often less is more. Simple and clear illustrations are better than busy slides. Colour and graphical elements should also be used sparingly. The point is to tell a story and deliver a message and not to overwhelm the listener with graphical information. 

Whether the graphical information and the images are gathered from internet searches or originally produced for the presentation, they must be closely related to the topic of the talk and should avoid distraction. A careful presenter will consider all types of listeners, including listeners with attention issues that might be easily distracted by superfluous information in the presentation, excessive use of colours, or fancy animation tricks. Small defects in the alignment of elements in a figure or in text can also be very distracting to some listeners.

\begin{figure}[htb]
\centering
\subfigure[Block Diagram \label{subfig:BlockDiagram}]{%
\includegraphics[scale=0.22]{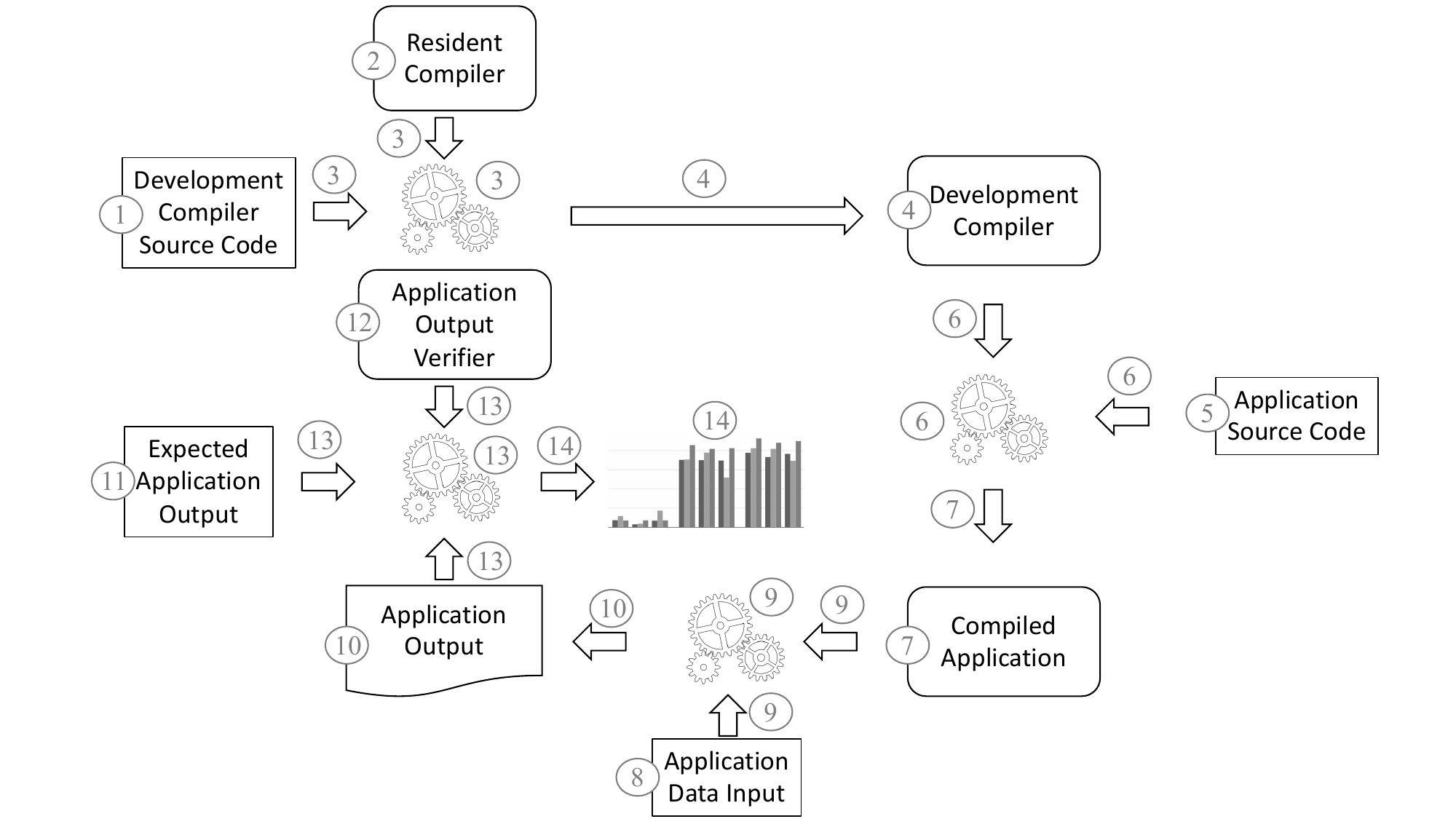} }
    \subfigure[Bulleted Text \label{subfig:BulletedText}]{%
    \includegraphics[scale=0.22]{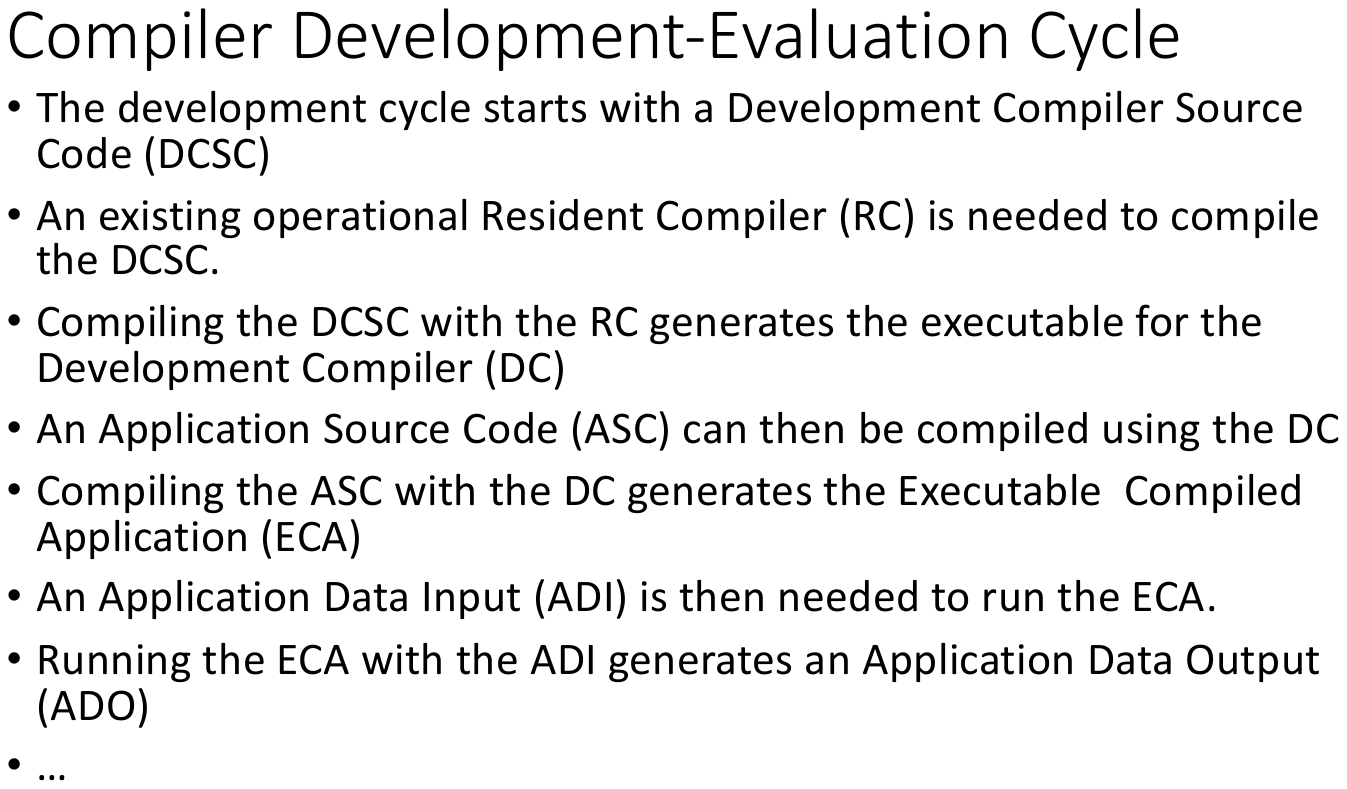} }
    \caption{\label{fig:vecAdd} Two versions of a single slide to explain the Compiler Development-Evaluation Cycle.}
\end{figure}

\subsection{Gradual Introduction of Complex Information}

Scientific presentations often include complex information that may take various forms. How this information is introduced will have a great influence in how effective the presentation will be and in how easy it will be to understand the information presented. A good mental model is to imagine that you are explaining a complex process to a person while standing in front of a white board. Suppose that you will illustrate this complex process using a block diagram. Most likely you will not silently draw a very complex diagram without saying a word and then start speaking. The most natural progress is for you to draw an initial block, label it, and explain its meaning, function, etc. Then you will draw a second element of your block diagram, explain what it is and how it connects with the first one, before introducing the next element. This is a natural interaction process because this is a very effective way to explain a complex process when you feel free to decide when you will present the graphical and the oral information in your explanation. However, when presenting from a set of slides, quite often, speakers will show a very complex diagram, graph, or a piece of program code all at once and then start talking about it. The natural reaction for many listeners is to feel overwhelmed by the excessive information and to tune out from the presentation.

\subsubsection{Presenting block diagrams or complex illustrations.}

As an example, consider the slide shown in Figure~\ref{subfig:BlockDiagram}. The numbers inside the circles would not appear in an actual slide for a presentation, they are added here to illustrate how the gradual presentation of complex information enhances understanding during an oral presentation. The title of the slide "Compiler Development-Evaluation Cycle" does not appear in  Figure~\ref{subfig:BlockDiagram} because it would have appeared, by itself, in the immediately precedent sign-posting slide. The oral presentation of the slide shown in Figure~\ref{subfig:BlockDiagram} would go as follows.

\begin{quote}
We have the source code for the development compiler \textcircled{1} and the host machine must have a resident compiler installed \textcircled{2}. Compiling the development compiler with the resident compiler \textcircled{3} produces the executable for the development compiler \textcircled{4}. Compiling an application program source code \textcircled{5} with the Development Compiler \textcircled{6} produces a compiled application \textcircled{7}. Then we need data input for this application program \textcircled{8}. Executing the compiled application with the application data input \textcircled{9} produces an Application Output \textcircled{10}. Now it is often necessary to not only verify that the output produced is correct, but also to collect performance data to assess the quality of the code generated by this version of the development compiler, thus we need an Expected Application Output \textcircled{11}. We also need a computer program, which can be as simple as the Linux program {\tt diff}, which we call the Application Output Verifier \textcircled{12}. Running the verifier with the produced output and the expected application output \textcircled{13} determines if the generated code produces correct outputs. Often, a developer is also interested in collecting performance data to assess the quality of the code generated \textcircled{14}.
\end{quote}

The example above is included here to contrast the gradual presentation of this complex block diagram, as illustrated by the text above with the help of the circled numbers, with an alternative presentation that shows the entire complex diagram at once. Figure~\ref{subfig:BulletedText} shows an anti-pattern of how the same information could appear in a "traditional" presentation with bulleted text. This anti-pattern illustrates various very ineffective techniques often used in slides. It has too many words,  with long items and it uses many acronyms. Understanding would be diminished and the quality of the presentation would suffer even further if all the content of this text-rich slide were to appear at once. If such a slide has to be used because there is not enough time to prepare a visual form for the information, at least the presenter should take care to introduce one item at a time in synchrony with the progress of the oral presentation. A better approach would be to make each bullet item less wordy and to have the items appearing one at a time.

There are some practical ways to create slides that progressively display a block diagram or a complex illustration using current technology:
\begin{itemize}
\item if you are drawing your own diagram from scratch, first make a draft on paper and then create the full diagram in a single slide using the software that you will use for your presentation. This is the last slide that your audience will see. Now either use animation technology to make the various elements appear in the sequence that you intend to present, or make as many copies of the same slide as needed, removing one element at a time. This last technique has the disadvantage that if you need to make a correction or modification in the future, you may have to repeat that modification into many slides.
\item if you are presenting a complex figure from another source that you cannot animate easily. You can rely on an old technique that used to be called "strip-tease" when projectors with overhead transparencies were commonly used. Add blank boxes to cover portions of the complex illustration and gradually remove these boxes.
\end{itemize}

\subsubsection{Presenting Numerical Results in Graphs.}

The gradual presentation of information also applies to the presentation of numerical results in graphical form. Way too often presenters introduce a complex graph with several groups of multiple colour columns, legends, scales on axis, names on axis, etc. And then the presenter immediately starts talking about high-level observations and claims about the data without giving the listener an opportunity to understand the data that is being presented. This is true for many types of data presentation, including line graphs, pie charts, etc. A more effective presentation will first present an empty graph with the axis, the scales, the names on the axis. At this point the presenter will explain what is being measured, what is the meaning of the legends, the scales, etc. If the graph contains multiple sets of columns, the speaker can then make the names of the sets appear and explain what is their meaning and significance, why they were chosen. Next the speaker may make the first column for the first set appear along with the entry in the legend only for that column, following by the introduction of the subsequent columns, one at a time. Once the entire legend is presented in this fashion, the speaker can then show the remainder of the graph and discuss the high-level observations and claims that are being made based on that graph.

Unfortunately the technology that we have available today for the creation of graphs is not advanced enough for the easy creation of such animated graphs. There are, however, two techniques that can be used to create such animated graphs; which one is chosen depends on how complex the graph is and how much time is available to create the final presentation. The first technique consists on copying and pasting a graph to the presentation as would be the case in a traditional presentation. Then each element of the graph can be drawn on top of this graph so that these individual elements can be animated. If the entire graph is redrawn, the originally pasted graph can be deleted from the slide. The alternative, often faster, technique consists in applying the strip-tease technique to the graph: add several blank boxes on top of the graph and then make these boxes disappear in synch with the oral explanation. When using this technique with graphs, for a good quality presentation, it is important to strive to make it not apparent to the audience because it is likely that several members of the audience will be distracted, or even irritated, if the temporary hiding of information is apparent in the slides.

\subsubsection{Presenting Complex Algorithms or Program Code.}

In Computing Science many presentations include complex algorithms or program code. Unfortunately, way too often a speaker will present a complex code or algorithm all at once, not taking time to explain the data inputs, the function of the code or algorithm, the data and control flow or the structure. There is no point in simply flashing a complex program code to an audience. If it is important that the audience understands the entire algorithm or code, a gradual presentation, similar to the case of the presentation of block diagrams or numerical results in graph form, is recommended. This gradual presentation can be produced either through animation or through the strip-tease technique. Again it is important to time the appearance of different elements in the program with the explanation that is given. In some situations the speaker wants to show a complex code for the purpose of focusing on only one aspect of the code. In such cases a speaker may start by showing the entire code, but then she should highlight, for example using different colours or producing a larger print of a portion of the code, the sections that are featured in the oral presentation.

\subsubsection{Including Equations in Slides.}

There are many parallels between the inclusion of program code, graphs with data, and equations in a presentation. Equations should only be included if they are necessary to improve understanding and to enable the speaker to make a point. If the equation is complex, it should not appear all at once. The strip-tease method of covering parts of the equation with boxes with the same colour of the slide background works well. Most importantly, all the elements and symbols of an equation must be explained to the level in which the group with the least background in the subject matter will understand. Audience members that are more knowledgeable will be forgiving and understanding. Moreover, explaining the notation and the meaning in the equations may reveal differences in the use of notation even amongst experts in a field and will enhance communication.

\subsubsection{Presentation of Textual Information.}

In general, the presentation of textual information should follow the same gradual-presentation principle: a text only appears in front of the audience when it is mentioned in the oral presentation. Textual information can be very useful but should be used thoughtfully. Long sentences and large paragraphs are difficult to parse while listening to someone speaking. Usually the audience has to make a choice between listening to the speaker or reading the text. Sometimes an effective use of text is to help the audience follow the line of an argument or the topics of a presentation. In this case each element on a list should have only a few words. Another effective use of text is when the speaker's accent may make difficult for some members of the audience to understand what is being said. In this case the speaker may purposefully include in the slides words that could be confusing to the audience.

Text is also very effective in signpost slides. A signpost slide contains only a few words and is used to signal a transition in the presentation. It helps both the speaker and the audience to keep the presentation organized because it marks the important transitions in a presentation.

\subsection{Hooking the Audience at the Start of a Presentation}

All too often most members of the audience have electronic devices with them. Even though most believe that they can mustitask between some activity in their electronic device and listening to the presentation, in reality their use of such devices greatly distracts from the speaker's message. However, most listeners will pay attention to the first few minutes of the presentation to decide if they should continue listening or not. Thus, the speaker must use these first few minutes to engage directly with the audience, to gain the attention of the audience, and to interest them to listen to the remainder of the talk. A presentation must have a good {\it hook} at the start. A good hook is one that the majority of the audience can relate to and that captures their interest. It can be a question, a joke, a story, an image, the posing of a problem or of a conundrum. Interested readers should search for examples of good presentation hooks. Here are two examples of unique hooks that have been used to great effect.

Several years ago I was invited by the IBM Centre for Advanced Studies in Toronto to come and give a talk about Transactional Memory, a microprocessor technology that most academics in the area knew had been invented by Maurice Herlihy and J. Eliot B. Moss at the University of Massachusetts~\cite{HerlihyISCA93}. At the time there was great interest at IBM about this technology because IBM was the first company to implement Hardware Transactional Memory in their Blue Gene Q machine~\cite{WangTC15}. In the process of researching for the talk, I discovered that in reality the technology had been concomitantly and independently invented by Janice M. Stone, Harold S. Stone, Philip Heidelberg, and  John J. Turek at IBM and Purdue University~\cite{StonePDT93}. However, Stone et al. called the technique ``The Oklahoma Update" because when a transaction is executed, either all the actions in the transaction are committed or none are. They were inspired by the song "All Er Nothing"  from the Oklahoma musical. For the hook for my presentation, once I was introduced by the host, I started playing the video for that song from the musical and I stood at the centre of the stage in the auditorium looking at the audience while the song played for almost five minutes. Many of the people in the audience were busy developers who had rushed from their cubicles or from other meetings to my presentation. They had their laptops in their lap and were busy with other tasks as the talk was announced. During those five minutes I slowly saw one laptop after the other closing down and the audience started looking at me. Once the song finished I asked ``Do you know the connection between Transactional Memory, IBM, and this song?" after a dramatic pause I said ``If you do not learn anything from this talk, I promise that you will learn the answer to this question." I had a very attentive audience for the rest of my talk because I had managed to hook them through curiosity in the first few minutes.

The second example is from a talk that I gave by invitation from the Transdisciplinary Advanced Studies Institute (IEAT) at the Federal University of Minas Gerais (UFMG) in Brazil. A central topic of my presentation was a discussion of an issue that has hindered experimental Computing Science for decades: the incorrect summarization of results and the derivation of research claims from such incorrect summarizations. The {\em Mercado Central} is a very well known public market in Belo Horizonte where the UFMG is located. I researched local news and found that {\em Dona Zelinha} had been a well-known cheese monger in that market. Thus, for the hook for my talk, I started by telling a story, illustrated with pictures, about visiting that market to buy some well-known local cheese and made the story about the confusion created when I tried to pay for two kinds of cheese using the average of the percentage discounts --- the very issue that is a problem in many Computing Science research papers. The audience was immediately hooked by a story that had lots of local context, and by the jest in which I told about the surprise of the cheese monger upon realizing that a university professor did not know how to correctly apply different discounts to separate items. Later on, {\em Dona Zelinha} appeared in the talk when I discussed the issues about summarization of data in Computing Science research, connecting back to the hook.

Most importantly, given the interactions between audience and presenters today, starting the presentation with an outline slide that spends the first three or four crucial minutes of the presentation stating what the content of the presentation will be is certainly a wasted opportunity. In a longer presentation where an outline can be helpful for the audience to follow the presentation, such an outline can be presented after the hooking introduction. This is a rare case where presenting the whole text of the outline at once in a single slide may be effective in most cases. 

\subsection{Delivering the Presentation}

When delivering the presentation, the most important concern of a speaker should be to engage with the audience and to take time to check that the audience is following the presentation. Eye contact with several audience members is important, therefore a speaker should not be constantly looking at the slides or at notes. The use of silence, pause, changes of tone are also very important elements of a well-delivered presentation. Many resources are available on the web with suggestions for making a good presentation. One that I find very useful is Julian Treasure's TED talk about how to use your voice as an instrument to engage your audience~\cite{TreasureTED13}. An inexperienced speaker should rehearse the whole talk enough times to feel confident on what to say and how to say it, but not so many times that the talk appears memorized. A rule of thumb that I use to advise beginner speakers is that an in-house talk should be rehearsed four times and a public talk to an external audience should be rehearsed up to seven times. However, the rehearsing should not take away from the goal of being engaged with the audience.

\subsection{Wrapping Up the Presentation}

The way a presentation is ended is almost as important as the way it is started. All too often a presentation ends with a busy slide with a bullet list of points. The conclusion is a moment to retell the presentation briefly and to emphasize the main messages. A good way to do this is to use the same graphic elements that were used to support the presentation earlier. At this point complex graphic elements can be introduced all at once as they have already been explained before. They can also be reduced in size so that several of them fit into a single slide. The idea is to use the human ability to recall graphic representation to emphasize the take-home points to the audience.

\subsection{The ``Elevator" Presentation}

Important exchanges between scientists occur in casual settings. While senior researchers are experienced with this type of interaction, for graduate students it is also important to be prepared to give an impromptu "elevator" presentation. The context for this kind of interaction is a casual hallway or coffee-break meeting with a more senior researcher in the same area.  When the conversation turns to the graduate student's research, the student should be able to give a brief presentation that focuses on the high-level formulation of the research and on the most important research problems or challenges being addressed. This is a different kind of presentation because there are no visual elements, no notes or slides. Making a good impression in such casual meetings can lead to many developments in a student's career. For instance, the senior researcher may be motivated to attend the student's talk later and is likely to pay more attention to it. Future placement opportunities may be also influenced by such casual meetings. A student should think about and practice multiple such short presentations that emphasize different aspects of the research as this kind of presentations may vary depending on the listener and the context of the meeting.

\section{Creating a Manuscript to Report Scientific Results}

Much has been written about effective creation of scientific manuscripts and also about the many ways in which authors may produce incorrect, misleading, or unclear presentation of results. Stephen Blackburn et al. discuss many ways in which a well-intentioned author can mislead a reader or misrepresent a result~\cite{BlackburnTOPLAS16}. They classify and categorize these shortcomings in the creation of scientific manuscript as ``sins" that are broadly divided into sins of exposition and sins of reasoning. The central idea is that errors may be committed not only by incorrectly presenting results, but also, and quite often, by incorrectly interpreting the data and deriving claims from that data. That paper was originated in work done for a few years in the Evaluate Collaboratory where several anti-patterns --- examples of ways in which scientists get things wrong --- were identified~\cite{EvaluateAntiPatterns}.

A section in Marie desJardins' article on how to succeed in graduate school also provides sound advice on how to write both a thesis and a conference or journal paper~\cite{desJardinsACMSM94}. Her key advice is that a paper must be relevant and must be well written. I would add to it that an important reflection when considering writing a paper is to reflect on what style of paper it is. What is the main purpose of the paper? Some typical main thrusts for papers in Computing Science are:
\begin{itemize}
\item presenting new theoretical results
\item presenting a new idea, new technique, or new algorithm
\item evaluating existing ideas and contrasting them
\item exploiting the use of a new technology and its impact on the state of the art
\item a reproduction study that carefully evaluates ideas previously published
\item surveying a research area, comparing and contrasting solutions
\end{itemize}

Often manuscripts will have a combination of the goals listed above. For example, the presentation of a new idea may include some performance evaluation, or a manuscript whose main goal is to evaluate and contrast different solutions to a problem may also introduce some new ideas. However,  
referees often will be reading a manuscript with the question ``What is this manuscript about?'' in their minds. Thus, it is important to organize the writing in such a way that it is easy for a reviewer to answer this question. When writing a manuscript, it is important to think as a fair-minded referee that is reading that manuscript for the first time. A classical writing by Alan Jay Smith provides a good summary of considerations that such a referee will make when assessing a manuscript~\cite{SmithComputer89}.
The Evaluate Collaboratory created a letter to Program Committee chairs advocating that more reproduction studies should be published~\cite{EvaluateLetter}. Unfortunately, it is still difficult to have even carefully performed and described reproduction studies accepted in conferences in Computing Science.

\subsection{Describing Experimental Methodology in a Manuscript}

All too often manuscripts that are reporting on experimental results fail to properly describe the experimental methodology used to produce the results. It only makes sense to present experimental results that show trends that would be observed again when the experiments are repeated in the future. It is even better if the trends would be the same if the evaluation were to be reproduced by different people in different machines. Thus, as a referee, I have several levels of confidence in the results presented in a performance-evaluation section of a manuscript:
\begin{enumerate}
\item Will similar trends in the data appear if the authors were to repeat the experiments on the same machine?
\item How about if others were to repeat the experimental evaluation on the same machine?
\item Would the results be leading to similar claims if the experiments were performed by the authors in different machines, versions of software systems, CPUs, and memory configurations?
\item How about if others were to reproduce the experiments in different systems?
\item For how long do we expect the claims to hold as computer systems evolve over time?
\end{enumerate}

Surprisingly, many manuscripts do not meet the even the most basic criteria of confidence in their experimental results. The description of the experiments in the manuscript does not offer indication that the authors themselves could reproduce the experiments in their own system. There are several simple guidelines that authors can follow to increase the confidence of a referee in an experimental evaluation:
\begin{inparaenum}[i)]
\item Clearly and concisely describe the machines and the software systems used for the performance evaluation, including version numbers, clock frequencies, amounts of memory and sizes of caches.
\item State the number of times each measurement was performed, what variations were observed between repetitions of the same experiment and how are these variations reported. Do they affect the confidence in the trends reported and used to state the claims?
\item State if dynamic voltage and frequency scaling was in operation in CPUs, GPUs and other accelerators when the experiments were run~\cite{BaoTACO16}. Most computer systems have frequency and voltage scaling automatically enabled. This scaling introduces external variations that can significantly affect the performance results. If any mitigation was done to reduce or eliminate the influence of scaling on the experimental results, such mitigations should be stated in the manuscript.
\item Since the discovery of the Spectre, Meltdown, and other side-channel vulnerabilities~\cite{KocherSP19,LipppUSENIX18}, many systems have implemented software mitigations to these vulnerabilities with significant impact on performance. If comparing performance across different systems, it is important to be consistent and either apply mitigations to all system or to none. In such case, when applicable, the manuscript should state the mitigation that was applied to each system.
\item State the order in which the experiments were performed. A good practice is to run through all experiments and all configurations once and then shuffle the order of the experiments before executing them next time and to continue in this fashion until the desired number of repetitions are completed. This shuffling technique prevents external variables, such as frequency and voltage scaling, from introducing bias towards a set of measurements. External variations are likely to appear as variations between multiple repetitions of the same measurement in this case. Thus, making it more important to report such variations in the presentation of the results.
\end{inparaenum}

\section{Measuring Performance}
\label{sec:perform}

John Ousterhout has produced an excellent article on many things that often go wrong during the collection of performance measurements and on the preparation of the results for presentation~\cite{OusterhoutCACM18}. This section is a combination of observations that we have made over many years in our research group and highlights of some of the excellent observations made by Ousterhout.

\subsection{Methodology  for Data Collection}

When performing experimental evaluation, computing scientists often work under the wrong assumption that the measurements could be easily repeated some time later if needed. Most of the time the very same experimental measurements cannot be repeated in the future because of frequent software system updates, hardware replacement, and changes to network configurations. A consequence of the lackadaisical attitude toward experimental data is that computing scientists have a tendency to be sloppy about data annotation and organization of experiments. Advising incoming students to think about experimental data recording the way some biologists would can be helpful. For instance, consider a biology researcher that is studying effects of different variables in the growth of some plants. The entire experiment may take several months to complete. If such a researcher were to mislabel data or haphazardly archive the data from the experiments, potentially there would only be an opportunity to repeat the experiment a year later. Thus, experimental biologists will be very careful about the collection, labeling, and archiving of experimental data. A responsible computing scientist should have the same level of care with the data. Practical suggestions include devising a naming scheme and a file structure for the storing of the data prior to starting the collection of data. Creating processes and scripts that automatically record all the information that will be needed at a later date together with the numerical results is also important. This information includes the computer specifications, versions of operating systems, compilers, network controllers, compilation flags, memory capacities, date and time of the experiment. As much as possible the transfer of values from the files where they are originally stored to a manuscript should also be automated to avoid clerical mistakes. A great advantage of an automated process for experimental evaluation is that often, while examining data from preliminary experiments, researchers will start asking questions about what would happen if changes were applied to the system. An experimentalist that has invested time upfront to automate the data collection process will be much more willing to regenerate the experimental data after changes than an experimentalist that has to labour for hours or days in order to generate the results for each change in the system. Making the data collection, as well as the generation of graphs and tables, easier, makes it less likely that potential alternative solutions be disregarded as not worth pursuing.

\subsection{Summarizing Results}

\begin{figure}[htb]
\centering
\subfigure[Synthetic Experimental Data \label{subfig:ExecutionTime}]{%
\includegraphics[scale=0.14]{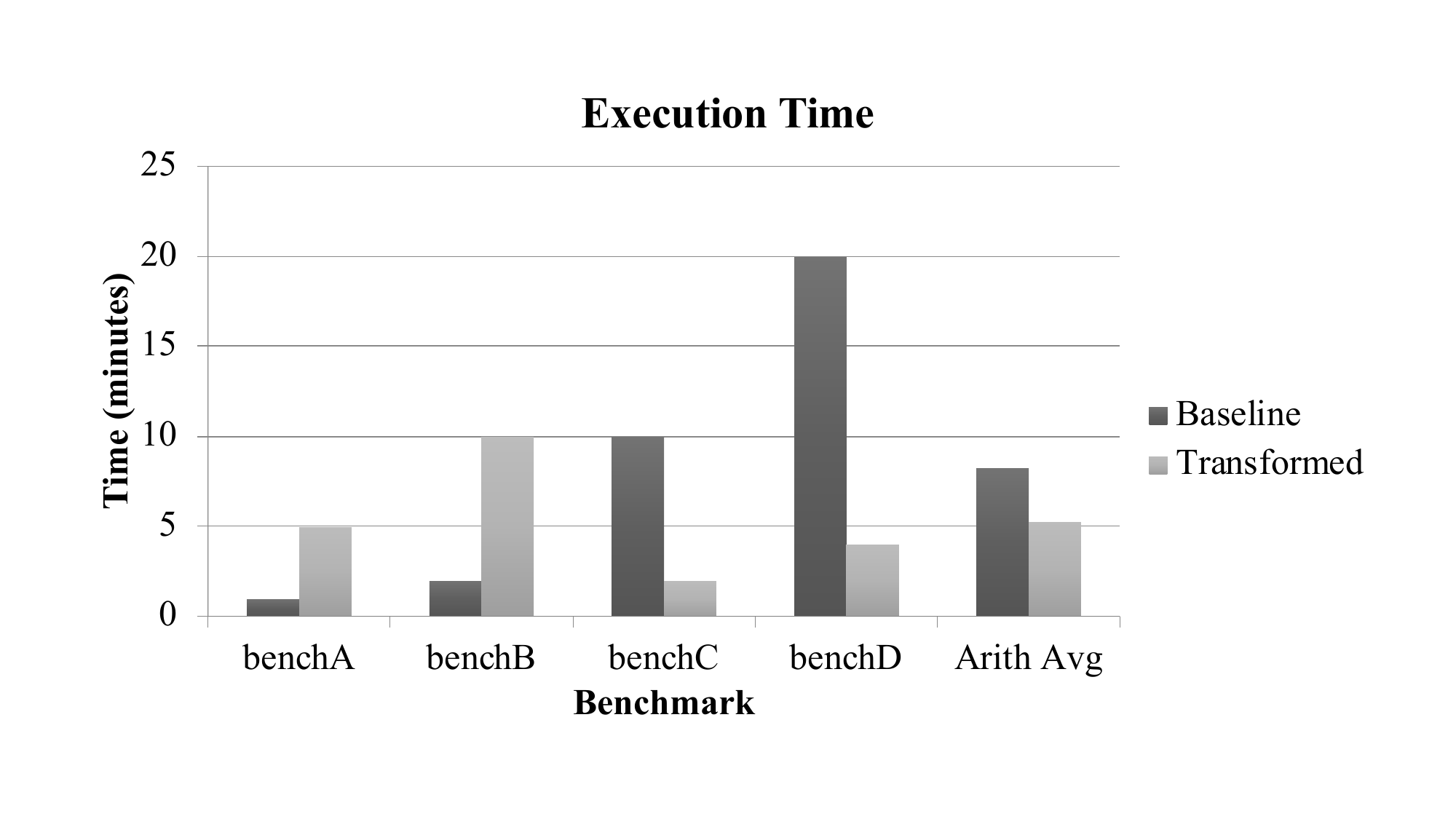} }
    \subfigure[Speedups  \label{subfig:Speedup}]{%
    \includegraphics[scale=0.14]{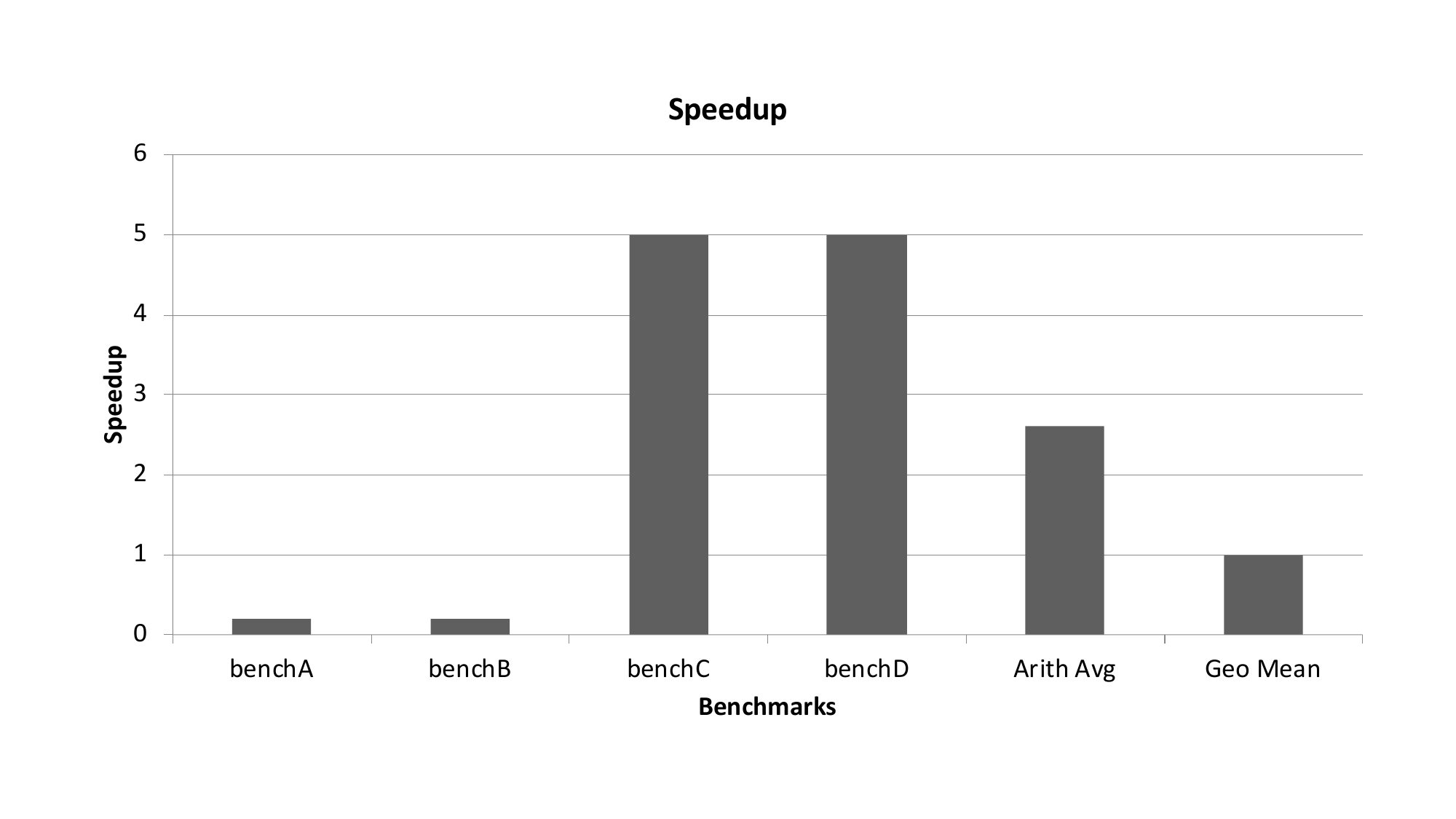} }
    \subfigure[Normalized Times \label{subfig:NormalizedTime}]{%
    \includegraphics[scale=0.14]{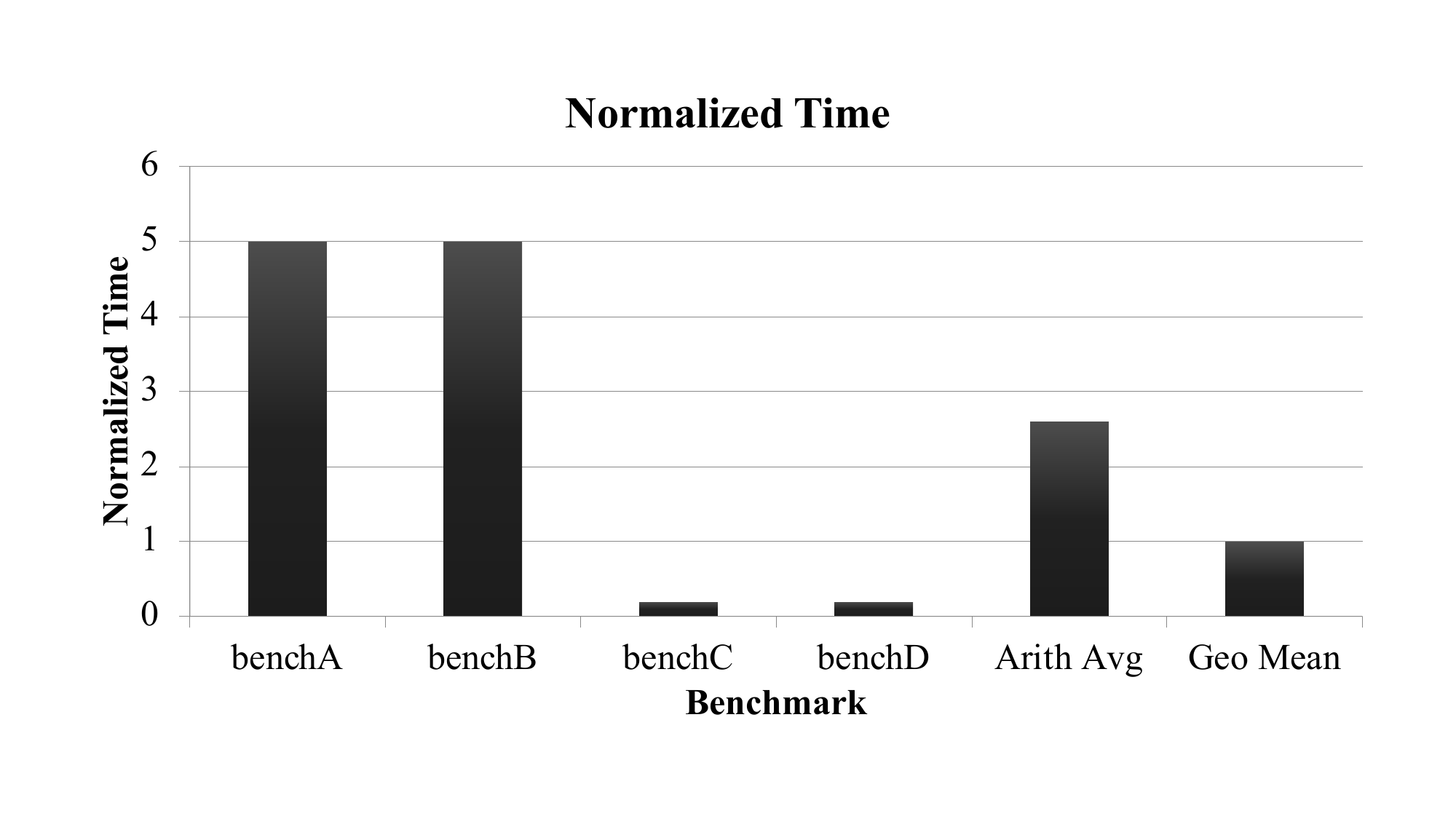} }
    \caption{\label{fig:vecAdd} Two ways to compare the results from an experimental measurement.}
\end{figure}

A surprisingly common error in the summarization of experimental data in computing science is the use of the arithmetic average to summarize normalized quantities --- the arithmetic average of percentages is a particular case of averaging normalized quantities that is specially baffling given that no shop keeper would allow such summarization to compute the final price of two items on sale for different discounts. Mathematically, this summarization technique is always wrong, and it is meaningless. Years ago Philip Fleming and John Wallace published a clear article explaining why such summarization is wrong~\cite{FlemingWallaceCACM86}. When the normalized values are close to each other, or when the base of the normalization is similar for all normalized values, the arithmetic average may approximate the correct summarization. Nonetheless, such cases do not justify its incorrect use.

However, when any of these conditions is not true, selecting this wrong summarization technique can lead to the opposite claims than what the data supports. To illustrate this point, years ago my student, Iain Ireland, created the synthetic data shown in Figure~\ref{subfig:ExecutionTime}. This data represent the execution time, measured in minutes, for four benchmarks in a baseline system and in a transformed system. Often, after generating such data, we are interested in the speedup of the transformed system over the baseline system which is computed as the ratio  $\frac{\mathrm{Time}_\mathrm{Baseline}}{\mathrm{Time}_\mathrm{Transformed}}$ shown in Figure~\ref{subfig:Speedup}. The arithmetic average of these speedups is 2.6 and thus we could conclude that the transformed system is, on average, 2.6 times faster than the baseline system. However, an equally acceptable presentation of the data would be to report the normalized time, which is computed as the ratio $\frac{\mathrm{Time}_\mathrm{Transformed}}{\mathrm{Time}_\mathrm{Baseline}}$, producing the results shown in Figure~\ref{subfig:NormalizedTime}. Thus, this arithmetic average would lead us to conclude that the transformed system takes, on average, 2.6 times longer to execute the programs, and therefore is 2.6 times slower than the baseline system, which is the exact opposite of the claim that the arithmetic average would lead us to make in Figure~\ref{subfig:Speedup}. Both are incorrect. If the desired comparison is for the latency to complete each individual benchmark, then a more suitable summarization is the geometric mean, which, in both cases would lead to the claim that the performance of the baseline and the transformed systems is the same. However, if the comparison should focus on the throughput, which is the total time required to execute all the benchmarks in each system, then comparing the arithmetic average of the original execution times, as shown in Figure~\ref{subfig:ExecutionTime} is correct and would lead us to conclude that the transformed system is 1.6 times faster than the baseline. However, in general, the summarization of rates should be done with great care as, in general, the geometric mean has no simple interpretation and can also be misleading for some experimental data~\cite{HoeflerBelliSC15}.

\subsection{Faith in Numbers and Estimations Instead of Experiments}

Two common problems described by Ousterhout are the excessive faith in numbers generated by a computer and the practice of estimating and guessing instead of measuring. The first issue relates to the fact that all too often researchers will collect numbers from experimental measurements and assume that the numbers are meaningful and are measuring what they intended to measure. Instead researchers should be skeptical about numbers produced by a computer and should examine all the programs, and the entire methodology, several times over, before starting making claims about the results.
A trivial way to spot issues is to take time to examine the raw data to spot possible problems with the experimental setup. Simply sorting the data in different ways and examining it can reveal significant outliers that might be indicative of an issue with the experimental framework. For instance, when measuring performance on GPUs a few years back, we noticed higher than expected variances in the measurements. Examining the raw experimental data we realized that the same programs were taking longer to execute when they were executed later in a given batch of experiments. We realized that we had not turned off automatic frequency scaling and the lower performance was caused by the GPUs getting hotter.

The second issue is related to the fact that while explaining unexpected results, too often authors make guesses and suppositions about the causes of these results but do not actually measure to confirm that their guessed culprit is indeed the cause of the observed results. A common situation is to make statements about the cache performance, the virtual memory paging system, the OS, or the branch predictor. Most systems have the capabilities that allow for a measurement of such effects. In my own experience, often when such a supposition is made, measurements can lead to much more insight about what is going on in the system and sometimes lead to further experiments that result in different claims about the system.

\subsection{Confirmation Bias}

In general, confirmation bias is defined as a tendency to treat new information as confirmation of our existing beliefs. In experimental computing science a common manifestation is in the different level of scrutiny for experiments that confirm or contradict our hypothesis. Suppose that we are measuring a transformation to a system that we designed and that we expect to improve the performance of the system. If we run a set of experiments and the results confirm our hypothesis that our transformation will improve the performance of the system, we are likely to readily accept those results as correct and the experiment as sound. However, if the results contradict our hypothesis and indicate that the transformed system is actually worse than the baseline system, we are likely to question the validity of the experiment and to carefully inspect each of the steps to find out what went wrong. This careful scrutiny should be applied in both cases.

\subsection{Hidden Learning in System Research}

A common problem in the development of computer systems is that often researchers have a scant amount of benchmarks to evaluate their system. Thus, it is quite common that the very same set of benchmarks with the very same set of data inputs will be used both during the system development and for its evaluation. The problem is that during development decisions are made based on preliminary evaluations of the system using that set of benchmarks. In many systems there are thresholds or constants that are set based on the discovery of a "sweet spot" using such benchmarks. If, after the development phase is over, the very same set of programs and data inputs are used to evaluate the system there is a high likelihood that the system is overfitted to those programs and the benefits to programs not yet seen is not as significant as advertised. We call this issue the "hidden-learning problem" of experimental computing science.

\subsection{Training vs Testing Data Set}

A similar issue occurs in performance evaluation that uses benchmarks that are often distributed with a very small set of data inputs. This limitation is particularly worrying for techniques that rely on profiling of previous executions of the program to make decisions. For a given class of computer programs, consider dense matrix multiplication and stencil computations, the behaviour of program is not affected by different numerical inputs. However, for other programs --- consider alpha-beta search, pathfinding, graph or tree based algorithms, applications based on hashing --- the data input can have a significant effect on the runtime behaviour of the program. In an effort to improve the availability of data inputs for benchmarks, the Alberta workloads for the industry-standard SPEC CPU2017 benchmark suite was developed~\cite{AmaralISPASS18}. The hope is that, with more data available, conscious researchers will be able to design sounder experimental methodologies to evaluate system innovation.

\section*{Concluding Remarks}

While the content of this article summarizes observations and practices collected over many years, the article was written after these practices were collected into the material taught in a graduate course at UFMG for the occasion of my visit sponsored by the IAET. There are two central themes of this article. The first is that effective communication occurs when the sender focuses on the receiver. This articles provides practical advice on how to prepare and deliver scientific presentations that are more effective at communicating scientific ideas and results. The second theme is that there are many practices in the generation and in the reporting of experimental results that make these results less reliable and that can lead to incorrect claims. An important reflection is how an academic discipline has become so lackadaisical about the reporting of experimental data. A possible cause is that the data is not worth much. To change this, we call upon researchers to define research questions that matter to them and to others, to strive to produce results that are relevant and informative to others and to carefully report these results. It is also important to become much more vigilant about careless experimental methodology or incorrect reporting of results and to point them out during the refereeing process as well as after publication.

\section*{Acknowledgements}

This article would not exist without the encouragement and support from Fernando Magno Quint\~ao Pereira. I am also thankful to the students at UFMG that enthusiastically participated in the course that I organized there and to the many students that were part of the research group in Alberta over many years. Special thanks to Artem Chikin and Matthew Gaudet for the careful reviewing of the manuscript. My visit to UFMG was supported by the Instituto de Estudos Avan\c{c}ados Transdisciplinares.
\bibliographystyle{acm}
\bibliography{local}
\end{document}